\documentclass[aps,pra,twocolumn,superscriptaddress,longbibliography]{revtex4-1}

\usepackage{natbib}
\usepackage{pgfplots}
\usepackage[breaklinks]{hyperref}
\usepackage{amssymb}
\usepackage{amsmath,amsthm}
\usepackage{amsfonts,dsfont}
\usepackage{graphicx}
\usepackage{mathtools}
\usepackage{tikz}
\usetikzlibrary{positioning}

\usepackage[normalem]{ulem}
\usepackage{color}
\usepackage{xcolor}

\newcommand{\de}{{\rm d}}

\begin{document}
\title{Homogeneous electron gas in arbitrary dimensions}

\author{Robert Schlesier}
\affiliation{Institut f\"ur Physik, Martin-Luther-Universit\"at
Halle-Wittenberg, 06120 Halle (Saale), Germany}

\author{Carlos L. Benavides-Riveros}
\affiliation{Institut f\"ur Physik, Martin-Luther-Universit\"at
Halle-Wittenberg, 06120 Halle (Saale), Germany}
\affiliation{NR-ISM, Division of Ultrafast Processes in Materials (FLASHit), Area della Ricerca di Roma 1, Via Salaria Km 29.3, I-00016 Monterotondo Scalo, Italy}

\author{Miguel A. L. Marques}
\affiliation{Institut f\"ur Physik, Martin-Luther-Universit\"at
Halle-Wittenberg, 06120 Halle (Saale), Germany}

\date{\today}

\begin{abstract}
  The homogeneous electron gas is one of the most studied model systems in condensed matter physics. It is also at the basis of the large majority of approximations to the functionals of density functional theory. As such, its exchange-correlation energy has been extensively studied, and is well-known for systems of 1, 2, and 3 dimensions. Here, we extend this model and compute the exchange and correlation energy, as a function of the Wigner-Seitz radius $r_s$, for arbitrary dimension  $D$. We find a very different behavior for reduced dimensional spaces ($D=1$ and 2), our three dimensional space, and for higher dimensions. In fact, for $D > 3$, the leading term of the correlation energy does not depend on the logarithm of $r_s$ (as for $D=3$), but instead scales polynomialy: $ -c_D /r_s^{\gamma_D}$, with the exponent $\gamma_D=(D-3)/(D-1)$. In the large-$D$ limit, the value of $c_D$ is found to depend linearly with the dimension. In this limit, we also find that the concepts of exchange and correlation merge, sharing a common $1/r_s$ dependence.
\end{abstract}

\maketitle

\section{Introduction}

The homo\-ge\-neous electron gas (HEG) is one of the most fundamental models of condensed mat\-ter theory~\cite{PeterReview}.  Despite its apparent simplicity, it has pla\-yed a crucial role in the development of electro\-nic structure theory for almost a century~\cite{Sommerfeld1928, Wigner1934, PhysRev.139.A796,PhysRevB.15.2819,PhysRevB.66.235116, PhysRevLett.105.086403,PhysRevLett.107.110402,PhysRevB.85.081103,JEP_2018__5__79_0}. In 1965 Kohn and Sham showed that the exchange and correlation energy of the HEG can be used to perform accurate many-body calculations for atoms, molecules and solids~\cite{PhysRev.140.A1133}. Since then, the HEG is one of the systems of choice to develop, improve and benchmark functionals in density functional theory (DFT)~\cite{Jones15}.

In the $D$-dimensional HEG model, an infinite uniform gas of electrons fills an infinite $D$-dimensional cube. The negative charge is neutralized by a uniform positive background. Not sur\-prisingly, simple metals (i.e., metal sodium) resemble quite well this pa\-ra\-digmatic system. While the exact kinetic energy and exchange energy were determined at the very beginning of quantum mechanics \cite{thomas_1927,dirac_1930},
the first analytical expression for the correlation energy had to wait until much later. This was obtained for high densities within the random-pha\-se approximation (RPA)~\cite{PhysRev.106.364}. Now we also have available highly accurate numerical values for the correlation energy of the HEG, for $D=1$~\cite{Helbig2011_032503}, 2~\cite{Attaccalite2002_256601}, and 3~\cite{PhysRevLett.45.566,Ortiz1994_1391}, from Monte-Carlo simulations.

For the spin-unpolarized HEG in $D = 1, 2$, and 3, the energy per
electron reads, as a function of $r_s$, the Wigner-Seitz radius
\cite{giuliani_vignale_2005}:
\begin{align}
\label{formula1}
\epsilon_D(r_s\rightarrow0) = \frac{a_D}{r_s^2} - \frac{b_D}{r_s} + c_D \ln r_s + \mathcal{O}(r_s^0) ,
\end{align}
where $a_D$, $b_D$ and $c_D$ are constants independent of $r_s$.
The first term in the right hand side of \eqref{formula1} is the non-interacting kinetic energy term while the second is the exchange one. The remaining terms represent the correlation energy.  The values of the constants in \eqref{formula1} are known for $D = 1, 2$, and 3~\cite{PeterReview}. The reconstruction of the series beyond the high-density regime is a fascinating field in itself with gaps still remaining in our knowledge (see, for a recent review, Ref.~\onlinecite{PeterReview}). 

One of the most basic and essential concepts in science, the dimensional parameter $D$ is usually investigated as it can yield remarkable insights into the physical $3D$ case~\cite{Ehrenfest,RevModPhys.54.407,
  doi:10.1063/1.457247,Gill3,Hendrych,PhysRevResearch.2.013217}.
There is a multitude of physical systems where one or two of the
physical dimensions are much smaller than the remaining ones. Such
systems can often be modeled as one and two dimensional (with the $3D$ form of the Poisson equation for the Coulomb interaction), reducing their complexity while keeping the most important qualitative features of $D=3$ \cite{Wagner2012,PhysRevA.93.021605,Schmidt2019}.  Indeed, by a meaningful dimensional crossover, insights from $3D$ systems are useful to study and develop exchange-correlation potentials for lower dimensional systems \cite{Dobson,Constantin}. Furthermore, and perhaps more importantly, reduced dimensions often exhibit notable physical properties. For example, in 1$D$ we find Luttinger physics~\cite{Voit_1995}, while the synthesis of graphene~\cite{Novoselov666} and related materials has opened the way to a myriad of novel physical effects in 2$D$~\cite{Liu2018}. In the large-$D$ limit, by increasing the degrees of freedom, the quantum world reduces to a classical one~\cite{Yaffe}. Finally, recent progress in the fabrication of artificial $2D$ materials paved the way for the artificial realization of non-integer dimensions. Indeed, fractal substrates  (e.g., Sierpi\'nski carpets of bulk Cu) confining electron gases have been already reported and the quantum states are found to exhibit also a fractal structure \cite{Kempes,PhysRevB.93.115428,PhysRevResearch.2.013044}.

 The main aim of this paper is to study the HEG in arbitrary dimensions. We assume that the integral form of the Coulomb interaction in $3D$ is valid at arbitrary dimensions. Note that this is the usual assumption for gases with dimensions larger than one. For $D=1$ the usual Coulomb interaction leads to divergence, so it is usually softened~\cite{PhysRevB.74.245427}. Other interactions have also been proposed to study the crossover between $2D$ and $3D$ systems~\cite{Constantin}. We consider the $D$-di\-men\-sional Cartesian space, that can be efficiently studied with the pla\-ne-wave basis set. Other, much more involved, types of geometries are important to study the strongly correlated (i.e., low-density) regime, as for instance hyperspheres. Unfortunately, due to the added complexity, that line of research has been limited to few electrons \cite{Gill2,GillLoos2014}. 
 
 We will pre\-sent analytic results for the $D$-dimensional HEG for the leading terms of the kinetic, exchange and correlation energies. Our main result is that for $D>3$ the ground-state energy of the HEG has a quite different behavior when compared to $D = 3$. Specifically, we will show that the energy for $D>3$ exhibits the following expansion in terms of the Wigner-Seitz radius $r_s$:
\begin{align}
\label{formula}
\epsilon_D(r_s\rightarrow0) = \frac{a_D}{r_s^2} - \frac{b_D}{r_s} + \frac{c_D}{r_s^{\gamma_D}} +   \mathcal{O}(r_s^0), 
\end{align}
where $\gamma_D = (D-3)/(D-1)$.
To obtain this result, we use the RPA that is known to
be exact in the limit of the dense gas. Incidentally, we note that the
usefulness of the RPA in materials science goes well beyond the study
of the HEG. In fact, RPA provides an excellent framework for producing
fully non-local exchange-correlation functionals, including long-range
van der Waals interactions~\cite{PhysRevB.61.16430,PhysRevB.64.195120,PhysRevLett.106.153003},
and static electronic correlations~\cite{Fuchs2012}. To go beyond the RPA and to obtain accurate results for the mid to low-density regime would require the use of Monte-Carlo techniques~\cite{doi:10.1063/5.0004608} or other high quality approaches such as the Singwi, Tosi, Land, and Sjölander method~\cite{PhysRev.176.589}.

This paper is organized as follows. We discuss first the kinetic and exchange energy for the $D$-di\-men\-sional gas. Although the results are both known we offer a new derivation for the exchange energy that is valid for integer and non-integer dimensions. We then compute the correlation energy for the HEG in arbitrary dimensions. The paper ends with a conclusion and two more technical Appendixes. Atomic units are used through\-out.

\section{Kinetic and exchange energy}

We use $k_{\pm}$ for the Fermi levels of the spin-up and spin-down channels, respectively. For convenience, we also define the quantity $k_{F}^{D}\equiv(k_{+}^{D}+k_{-}^{D})/2$. The relations $k_{\uparrow,\downarrow} \equiv k_{\pm}/k_{F}=(1\pm\xi)^{1/D}$ determine $\xi$, the system's spin polarisation, lying between $0$ and $1$. The Wigner-Seitz radius is written as a function of the uniform density $N/\Omega$ in the usual way: 
\begin{align}
\frac{\Gamma(\frac{D}{2}+1)}{\pi^{D/2}}\frac{1}{r^D_{s}}
= \frac{N}{\Omega}\,.
\label{radius}
\end{align}
Here $\Gamma$ is the gamma function, $N$ is the number of electrons and $\Omega$ the volume occupied by the electrons.

At any dimension, the one-electron orbitals are plane waves and the ground-state energy of the HEG can be obtained by perturbation methods yielding:
\begin{align}
\varepsilon_D(r_s,\xi) = \varepsilon^t_{D}(r_s,\xi) +
\varepsilon^x_{D} (r_s,\xi) + \varepsilon^c_{D}(r_s,\xi),
\end{align}
where the noninteracting kinetic energy $\epsilon^t_{D}(r_s,\xi)$ and the exchange energy $\epsilon^x_{D} (r_s,\xi)$ are the zeroth- and first-order terms of the expansion. The correlation energy $\epsilon^c_{D}(r_s,\xi)$ is computed from all higher orders.


The calculation of the kinetic energy for the HEG is a text-book problem and we present only the result. By integrating the energy contribution of each electron along the Fermi sphere one gets the following expression~\cite{PeterReview}:
\begin{align}
\varepsilon^t_{D} (r_s,\xi)=\dfrac{\alpha_{D}^{2}D}{{2}(D+2)}
\frac{\Upsilon_2(\xi)}{r_{s}^{2}},
\end{align} 
where $\alpha_{D} = 2^{(D-1)/D}\Gamma\left(D/2+1\right)^{2/D}$ and $$
\Upsilon_n(\xi) = \frac12 [(1+\xi)^{(D+n)/D}+(1-\xi)^{(D+n)/D}]
$$
is the spin-scaling function. The first prefactor for the spin-un\-po\-la\-ri\-zed case in Eq.~\eqref{formula1} and \eqref{formula} is thus $a_D = \alpha_{D}^{2} D / 2(D+2)$. In the large-$D$ limit (which is easily achieved by using the Stirling's formula) we have for this function a quadratic scaling with the dimension
\begin{align}
    a_{D\rightarrow\infty} = \frac{D^2}{2e^2} + 
    \frac{\ln(\pi) + \ln (D)}{e^2} D + \mathcal{O}(D^0) .
\end{align}
Here $e$ is Euler's constant.

The standard calculation of the exchange energy can also be easily generalized to arbitrary dimensions. Indeed, the exchange energy per particle
reads
\begin{align}
\varepsilon^x_{D} &=-\dfrac{\Omega}{2N}\sum\limits_{i\in\{+,-\}}\int^{k_{i}}\dfrac{\de^{D}p_{1}}{(2\pi)^{D}}\dfrac{\de^{D}p_{2}}{(2\pi)^{D}}U(\textbf{p}_{1}-\textbf{p}_{2}) ,
\label{exchange}
\end{align}
where $U(\textbf{q})$ is the $D$-dimensional Fourier transform of the Coulomb potential. All angular integrals in the expression \eqref{exchange} are straightforward except the one for the angle between the momenta $\textbf{p}_{1}$ and $\textbf{p}_{2}$. The integrals within are symmetric under the interchange of $\textbf{p}_1$ and $\textbf{p}_{2}$ which can be used to factor out one of the integrals over the magnitude of the momenta. After these steps one arrives at the expression
\begin{align}
\varepsilon^x_{D}&=-\dfrac{\Omega}{N}\dfrac{D}{D+1}\dfrac{(k_{+}^{D+1}+k_{-}^{D+1})}{\pi(4\pi)^{\frac{D}{2}}\Gamma(\frac{D}{2}+1)} I\left(\frac{D-1}{2}\right),
\end{align}
where we introduced the integral
\begin{align}
  I(a)&=\int\limits_{0}^{1}\int\limits_{-1}^{1}\dfrac{\eta^{2a}(1-u^{2})^{a-1}}{(1+\eta^{2}-2\eta u)^{a}}du d\eta.
  \label{eq:intI}
\end{align}
We were unable to evaluate analytically this integral for arbitrary values of $a$. However, we use a special relation with respect to the variable $\eta$ to construct a recursive relation. In the Appendix~\ref{app1} we perform this calculation in full detail, leading to the recursive formula
\begin{align}
I(a)-I(a+1)=\frac{1}{a(a+1)}.
\end{align}
In the special cases $a=1/2,1$ the integral~\eqref{eq:intI} can be performed analytically, resulting in $I(a)=1/a$ for all integral dimensions. Furthermore one can evaluate the integral numerically on the interval $a\in (0,1]$, leading to the result $I(a)= 1/a$ for all values of $a$. Now one can substitute the Fermi momenta with the corresponding expressions depending on the spin polarization. One eventually obtains the result of Ref.~\cite{glasser} for arbitrary integer and non-integer dimensions as well:
\begin{align}
\label{eqexchange}
\varepsilon^x_{D} (r_s,\xi)=-\dfrac{2\alpha_{D}D}{\pi(D^{2}-1)}
\frac{\Upsilon_1(\xi)}{r_{s}}.
\end{align}
Thus for the spin-un\-po\-la\-ri\-zed case in Eq.~\eqref{formula1} we have $b_D = 2 \alpha_{D} D / \pi (D^2-1)$.
In the large-$D$ limit this term goes to a constant value:
\begin{equation}
  b_{D\rightarrow\infty} = \frac{2}{e\pi}  + \mathcal{O}(1/D) .
\end{equation}
Notice that from Eq.~\eqref{eqexchange} one eventually deduces that the exchange energy scales as $(N/\Omega)^{1/D+1}$, with the same exponent of the Lieb-Oxford bound for the indirect Coulomb energy in  arbitrary dimensions \cite{PhysRevLett.102.206406,doi:10.1002/qua.25224}.


\section{Correlation energy}

The high-density correlation energy for the two- and three-dimensional HEG is well understood~\cite{PeterReview}. For instance, for $D=3$ the correlation energy has the following expansion~\cite{PhysRevB.84.033103}:
\begin{align}
\varepsilon^c_{3} (r_s,\xi) = \sum_j[\lambda_j(\xi) \ln(r_s) +\omega_j(\xi) ] r_s^j
\end{align}

For arbitrary dimensions in the high-density regime, we apply the known resummation technique for $D = 3$ and generalize accordingly to arbitrary dimensions. We follow closely the classical work of Gell-Mann and Brueckner~\cite{PhysRev.106.364}. Since the change of dimension does not modify the topology of the Feynmann diagrams, the sum of all the ring diagrams of the same order $(n)$ yields the usual form known from the RPA:
\begin{align}
\label{energyn}
E^{(n)}_{c}=(-1)^{n+1}\frac{\Omega}{2} \int \frac{d^{D}q}{(2\pi)^{D}} \left[\frac{U(q)}{(2\pi)^D}\right]^{n} A_{n}(\textbf{q}),
\end{align}
where the functions 
\begin{align}
\label{an}
A_{n}(\textbf{q})&=\dfrac{1}{n}\prod\limits_{k=1}^{n}\left(2\int\limits_{-\infty}^{\infty}dt_{k}F_{\textbf{q}}(t_{k})\right)\delta\left(\sum\limits_{k=1}^{n}t_{k}\right)
\end{align}
integrate the propagators
$F_{\textbf{q}}(t)=\frac{1}{2}\int d^{D}p \,e^{-|t|(q^{2}/2+\textbf{p}\cdot\textbf{q})}$.
The $D$-dimension Fourier transformed Coulomb potential is 
\begin{align}
U(q) = \dfrac{(4\pi)^{\frac{D-1}{2}}}{q^{D-1}}\Gamma\left(\frac{D-1}{2}\right).
\end{align}
Notice that we get in Eq.~\eqref{energyn} $n$ copies of $U(q)$ since we need $n$ interactions to connect the corresponding fermion loops and $n$ copies of $(2\pi)^{-D}$ from the momentum integrals over the fermion loops. The oscillating sign comes from the fact that every fermion loop comes with a minus. In the propagator $F_{\textbf{q}}(t)$, the momentum integrals are performed for the regions $A_{i}=(|\textbf{p}|<k_{i}) \cap (|\textbf{p}+\textbf{q}|>k_{i})$.

Scaling the momenta $\textbf{q}\rightarrow k_{F}^{D}\textbf{q}$, using the relation $k_{F}=\alpha_{D}/r_{s}$, and the definition of the Wigner-Seitz radius one eventually obtains the contribution of the $n$-th order ring diagram to the total energy per particle:
\begin{align}
\varepsilon^{(n)}_{c}&=-\left[\dfrac{\Gamma(\frac{D}{2}+1)\Gamma^{2}(\frac{D-1}{2})}{4\pi^{\frac{3D}{2}+1}}\right]\int\dfrac{d^{D}q}{q^{2D-3}}\nonumber\\
&\int\limits_{-\infty}^{\infty}du\frac{(-1)^{n}}{2\pi n}[Q_{\textbf{q}}(u)]^{n}\left(\dfrac{\Gamma(\frac{D-1}{2})r_{s}}{q^{D-1}\alpha_{D}\pi^{\frac{D+1}{2}}}\right)^{n-2}
\label{B}
\end{align}
with $Q_{\textbf{q}}(u) = \int dt F_{\textbf{q}}(t)e^{iuqt}$. In the RPA, the total correlation energy per particle amounts naturally to the sum of all these contributions. At this stage one can show that the individual ring diagram contributions of order greater or equal than two diverge. The divergence occurs at low momentum $q$. If one first focus on the low momentum behaviour of $Q_{\textbf{q}}(u)$ it turns out that this quantity is independent of $\textbf{q}$. Thus one can pull this factor out of the momentum integral if one stays close to the lower limit of the integration. After expanding the integral measure one arrives at an integrand which is proportional to $1/[q^{D-2}(q^{D-1})^{n-2}]$. Its integral diverges at the lower limit for all $D>2 \wedge n\geq 2$ and $D\geq 2 \wedge n>2$.

Since we are interested in high densities (i.e. $r_s\rightarrow 0$) the only relevant parts are the low momentum domain of the $q\ll1$ integral. We then drop all terms with $q$ raised to a higher power. After these approximations are performed we reach a relatively simple expression for $Q_{\textbf{q}}(u)$:
\begin{align}
Q_{\textbf{q}}(u)&\approx\dfrac{4\pi^{\frac{D-1}{2}}}{\Gamma(\frac{D-1}{2})}R_{\frac{D-1}{2}}(u,\xi)
\end{align}
where 
\begin{align}
&R_{\frac{D-1}{2}}(u,\xi)=\dfrac{1}{4u^{2}}\dfrac{\Gamma(\frac{3}{2})\Gamma(\frac{D-1}{2})}
{\Gamma(D/2 +1)}
\nonumber \\ & 
\left\{(1+\xi) Y_{\frac{D-1}{2}}\left[\frac{u}{(1+\xi)^{\frac1{D}}}\right]+(1-\xi)Y_{\frac{D-1}{2}}
\left[\frac{u}{(1-\xi)^{\frac1{D}}}\right]
\right\}.
\nonumber
\end{align}
Here $Y_{a}(z)= {}_2 F_{1}\left(1,3/2;a+3/2;\frac{-1}{z^{2}}\right)$
is the hypergeometric function. One can now perform the angular part of the $q$ integration and evaluate the series with the same type of Gell-Mann and Brueckner's convergence argument \cite{PhysRev.106.364}.

We can group the results for the RPA correlation energy in three  different cases (see Appendix~\ref{sec:nonV}):
 \renewcommand{\labelenumi}{(\roman{enumi})}
\begin{enumerate}
\item When $D=1 \vee D=2$ the leading contribution to the correlation energy is a constant term.

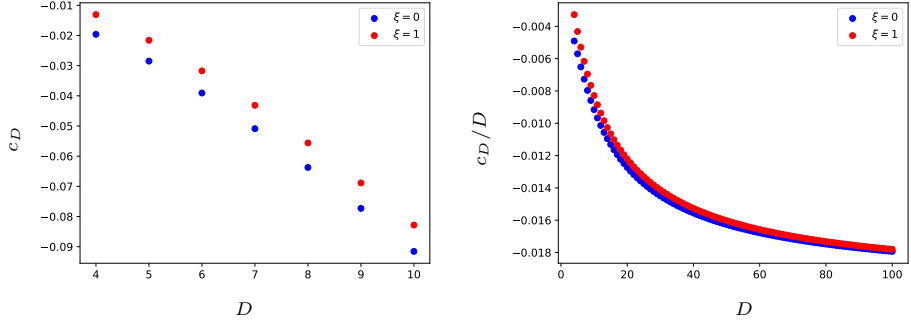
\begin{figure*}[t]
\begin{tikzpicture}
 \node (img) {
   \includegraphics[width=6cm]{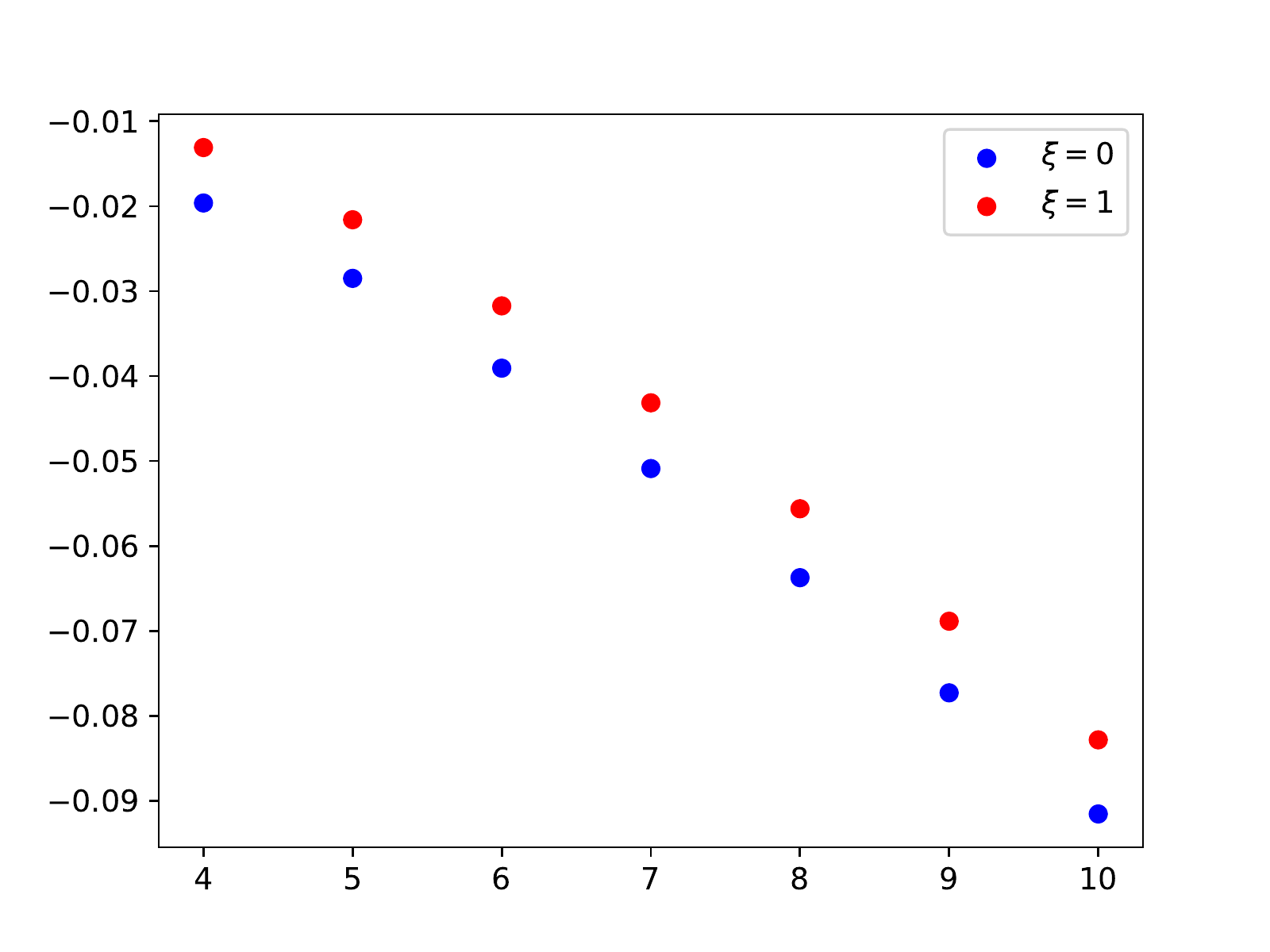}\hspace{0.25cm}
   \includegraphics[width=6cm]{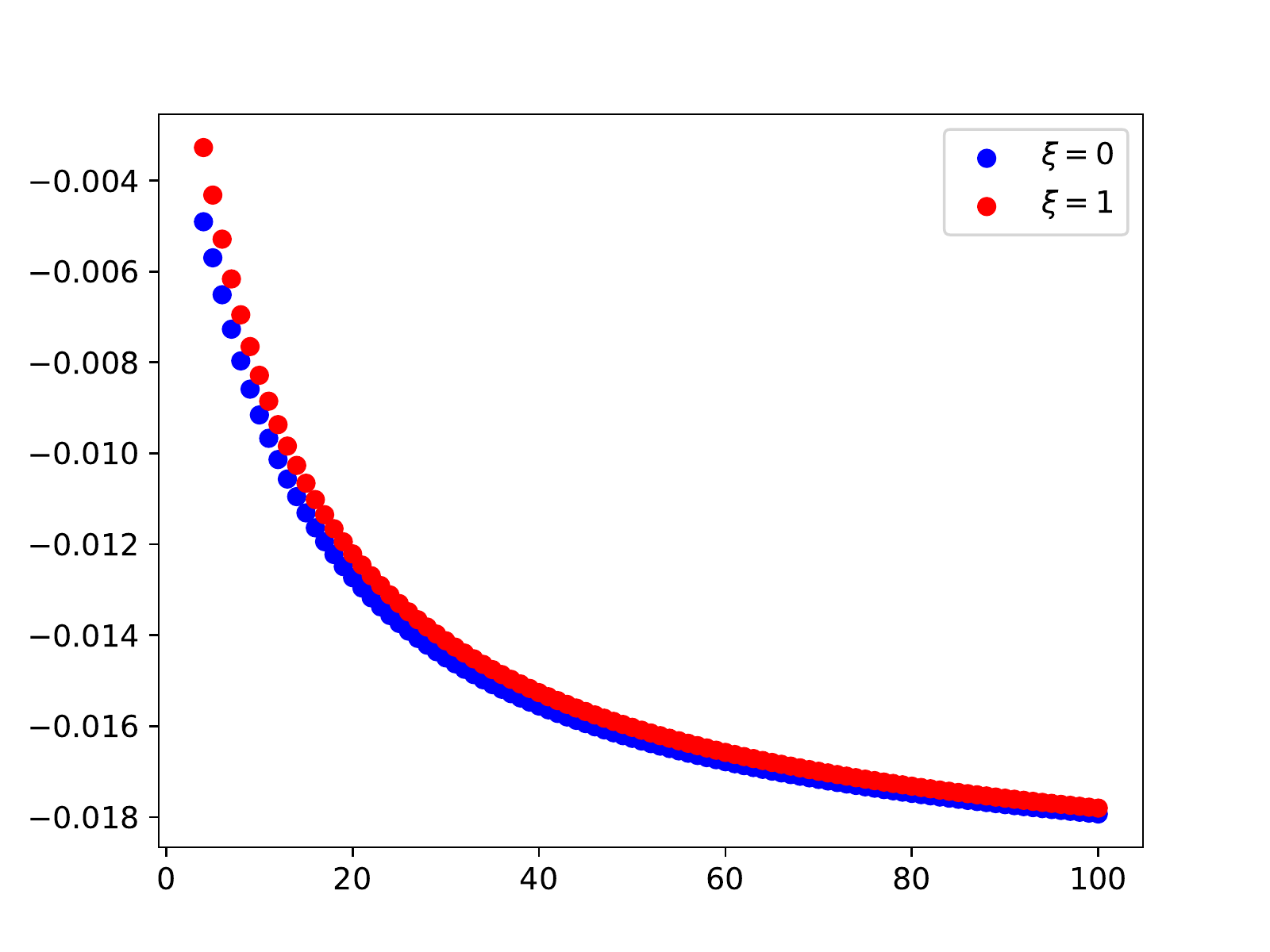}
 };
 \node[left=of img, node distance=0cm, anchor=center, xshift=1cm,yshift=-0.1cm,font=\color{black}] {\rotatebox{90}{\scriptsize$c_D$}};
 \node[left=of img, node distance=0cm, anchor=center, xshift=7.2cm,yshift=-0.1cm,font=\color{black}] {\rotatebox{90}{\scriptsize$c_D/D$}};
 \node[below=of img, node distance=0cm, xshift=-3.25cm, yshift=1.2cm,font=\color{black}] {\scriptsize$D$};
  \node[below=of img, node distance=0cm, xshift=3.4cm, yshift=1.2cm,font=\color{black}] {\scriptsize$D$};
 \end{tikzpicture}
\caption{Coefficient $c_D$ and $c_D/D$ as a function of the dimension $D$ for the HEG in the unpolarized and totally polarized limits.}
\label{fig1}
\end{figure*}

\item When $D = 3$ the leading term is the logarithmic correction known since 1950~\cite{Macke}: 
\begin{align} 
\varepsilon^{c}_3(r_s,\xi) &= \frac{1-\ln (2)}{\pi^2} \Upsilon_c(\xi)
\ln (r_s) +\mathcal{O}(r_s^0).
\end{align}
where the spin-scaling function for the correlation energy is
\begin{multline}
  \Upsilon_c(\xi)=\frac{1}{2}+\frac{(1-\xi^2)^{1/3} [(1+\xi)^{1/3}+(1-\xi)^{1/3}]}{4[1-\ln(2)]}
  \\
  - \frac{1}{4[1-\ln(2)]}
  \ln\left\{\frac{\left[(1+\xi)^{1/3}+(1-\xi)^{1/3}\right]^2}
  {(1+\xi)^{\frac{1+\xi}3}(1-\xi)^{\frac{1-\xi}3}}\right\}.
\end{multline}

\item For $D > 3$ the contribution is 
\begin{multline}
\varepsilon^{c}_D(r_s,\xi) = \\ -\dfrac{2D\Sigma_{D}}{\pi^{3}(D-1)}\left(\frac{\alpha_{D}\pi}{4r_{s}}\right)^{\frac{D-3}{D-1}}\int\limits_{-\infty}^{\infty}du\left[R_{\frac{D-1}{2}}(u,\xi)\right]^{\frac{D+1}{D-1}} \\
+\dfrac{D}{\pi^{3}(D-3)}\int\limits_{-\infty}^{\infty}du\left[R_{\frac{D-1}{2}}(u,\xi)\right]^{2}+ \delta_D,
\label{eqfinal}
\end{multline}
where $\Sigma_{D}$ is a special $D$-depend series (see Appendix~\ref{sec:nonV})
and $\delta_D$ regulates the full approximation such that the second order energy contribution is exact. Remarkably, the constant term in  Eq.~\eqref{eqfinal} does not diverge as $D\rightarrow 3$ from above, since $\delta_D$ exhibits the same divergence with the opposite sign.

Our result then  shows that for $D > 3$ the correlation behaves as 
\begin{align} 
\label{scaling}
\varepsilon^{c}_D(r_s,\xi) \sim r^{(3-D)/(D-1)}_{s}.
\end{align}
\end{enumerate}

\begin{table}[t]
 \centering      
{
\begin{tabular}{c|c c c c  c c c c c c c}
$c_D$  & 4 & 5 & 6 & 7 & 8 & 9  \\ \hline 
$\xi=0$ & $ -0.0196$ & $-0.0285$ & $-0.0391$ & $-0.0509$ & $-0.0638$ & $-0.0773$  \\
$\xi=\pm1$ & $-0.0131$ & $-0.0216$ & $-0.0318$ & $-0.0432$ & $-0.0556$ & $-0.0689$  \\
\end{tabular}}
\caption{Value of the coefficients $c_D$ for different integer dimensions $D$ in the unpolarized and totally polarized limits.}
\label{table:litioON678} 
\end{table}

We obtained numerically the prefactors $c_D$ (see Figure \ref{fig1}). In Table \ref{table:litioON678} we give explicitly the values up to $D=9$. Notice that they are all negative, as expected for a correlation energy. In the large-$D$ limit they scale linearly with the dimension: $c_{D\rightarrow\infty} = -D/2e\pi^2 + \mathcal{O}(D^0)$. Finally, from Eq.~\eqref{scaling} one obtains that in leading order the total correlation energy scales with the density as $(N/\Omega)^{(D^2-3)/(D(D-1))}$.

For the sake of comparison, we can now evaluate the equilibrium density that minimizes the energy of the HEG. Interestingly, the larger the dimension the larger the equilibrium $r_s$. For example, using only the exchange energy we obtain the equilibrium $r_s$ of 4.82337
($D=3$), 9.34001 ($D=4$), and 15.1596 ($D=5$). Including the RPA
correlation this values are reduced to 3.82865 ($D=3$),
8.73997 ($D=4$), and 13.3068 ($D=5$). This means that the
large-dimension HEG is stable for very low density gases. Recall that $r_s$ and the density are connected by the definition \eqref{radius}. Since the dimensional factor $\Gamma(D/2 +1) /\pi^{D/2}$ increases rapidly with the dimension, one can see that the equilibrium density decreases with the dimension but a much slower pace than the increase of $r_s$. 

\section{Conclusion and outlook} 

Recent progress in the physical realization of non-integer dimensions has stimulated the study of electronic gases in exotic dimensions, beyond the three-dimensional world we are \textit{a priori} used to. In this paper we studied the HEG in such situations, with special emphasis in the leading orders of the correlation energy. 

From our results, we can extract some interesting algebraic properties of the correlation energy of the $D$-dimensional HEG. First, our physical world, with $D=3$, stands out as completely different from either the reduced dimensions (1 or 2) or higher dimensions. For dimensions greater than $3$, the leading dependence on $r_s$ changes with the dimensionality, in contrast to the kinetic and exchange parts. Finally, for large dimensions the correlation energy goes as $1/r_{s}$ which is the same dependence as exchange. This means that for higher dimensions, the HEG no longer becomes weakly correlated for large densities, but is equally correlated in the whole range of densities.
We believe that this work on arbitrary dimensions can shed some light
into the more far-reaching problems of the correlation energy for the
HEG in fractional dimensions \cite{Kempes,PhysRevB.100.155135,PhysRevResearch.2.013044,PhysRevB.97.195101,PhysRevB.98.205116} and  its quantum-information properties \cite{PhysRevLett.112.150501,C7CP01137G, PhysRevA.95.032507,PhysRevLett.71.1291,2020arXiv200600961D}. This can also be a point of departure to develop a more coherent and unified dimensional approach to new exchange-correlation functionals within DFT. Indeed, for the cases of $2D$ and $3D$ such a dimensional unification has already proved to be quite successful \cite{Constantin}. 

\begin{acknowledgements}
We thank Peter Gill for helpful discussions.
\end{acknowledgements}

\onecolumngrid

\appendix

\section{Calculation of the recursive relation for the integral $I(a)$}\label{app1}

Let us define the auxiliary function $f(c,b,u)$, reading:
\begin{align}
\label{eq1}
f(c,b,u)=\int\limits_{0}^{1}\dfrac{\eta^{b}}{(1+\eta^{2}-2\eta u)^{c}}d\eta.
\end{align}
One can easily verify that this function satisfies the relation:
\begin{align}
f(c,b,u)= \frac{-(2-2u)^{1-c}+(b-1)f(c,b-2,u)+(c-b)2uf(c,b-1,u)}{2c-b-1}.
\end{align}
It turns out to be useful if one first obtains the antiderivative of $f$ (as a function of $u$) which is: $F(c,b,u)= f(c-1,b-1,u)/(2(c-1))$ and  its derivative: $f'(c,b,u)=2cf(c+1,b+1,u)$ before one starts with the main calculation. Now, choosing  $b=2a$ and $c=a$, Eq.~\eqref{eq1} leads to the following expression:
\begin{align}
I(a)=2^{1-a}\int\limits_{-1}^{1}(1+u)^{a-1}du+(1-2a)\int\limits_{-1}^{1}(1-u^{2})^{a-1}f(a,2a-2,u)du+2a\int\limits_{-1}^{1}(1-u^{2})^{a-1}uf(a,2a-1,u)du.
\label{2}
\end{align}
The first term on the right hand side is an easy integration giving $2/a$. Consider now the third term on the right hand side. We can partially integrate this term and under the condition $a>1$, the boundary term vanishes. This results in:
\begin{align*}
-2a\int\limits_{-1}^{1}\frac{d}{du}[(1-u^{2})^{a-1}u]F(a,2a-1,u)du=-\dfrac{a}{(a-1)}I(a-1)-\dfrac{a(1-2a)}{(a-1)}\int\limits_{-1}^{1}(1-u^{2})^{a-2}u^{2}f(a-1,2a-2,u)du.
\end{align*}
Let us now come to the second term on the right hand side of \eqref{2}. Here we partially integrate again. The boundary terms vanishes also for $a>1$. As before, we integrate $f$ and differentiate the rest of the integrand to obtain the following expression:
\begin{align}
-(1-2a)\int\limits_{-1}^{1}\dfrac{d}{du}[(1-u^{2})^{a-1}]F(a,2a-2,u)du
=(1-2a)\int\limits_{-1}^{1}(1-u^{2})^{a-2}uf(a-1,2a-3,u)du.
\end{align}
Inserting these three results in the right hand side of Eq.~\eqref{2} we obtain:
\begin{align}
I(a)=\dfrac{2}{a}+\dfrac{(1-2a)L(a)-aI(a-1)}{a-1}.
\end{align}
with 
$ L(a)= \int\limits_{-1}^{1}(1-u^{2})^{a-2}\left[(a-1)uf(a-1,2a-3,u)-au^{2}f(a-1,2a-2,u)\right]du$.

Now we use the relation \eqref{eq1} again, but this time inserting $c=a-1$ and $b=2a-1$, obtaining:
\begin{align}
(2-2u)^{2-a}-2f(a-1,2a-1,u)&=2(a-1)f(a-1,2a-3,u)-2auf(a-1,2a-2,u)
\label{3} . 
\end{align}
If we multiply \eqref{3} with $u(1-u^{2})^{a-2}$ and integrate over $u$ from $-1$ to $1$ we can see that the right hand side is equal to $2L(a)$. After manipulating the left hand side (partial integration of the second term, in this case the we differentiate the function f and integrate the rest of the integrand) we get $L(a)=-I(a)+ (a-2)/a(a-1)$. Finally, we insert the relation for $L(a)$ in \eqref{3} and obtain:
\begin{align} 
I(a)=\dfrac{1}{(a-1)^{2}}-\dfrac{aI(a-1)+(1-2a)I(a)}{a-1}\,\, \Rightarrow \,\, I(a-1)-I(a)=\dfrac{1}{a(a-1)}\,\, \Rightarrow \,\, I(a)-I(a+1)=\dfrac{1}{a(a+1)}.
\end{align}
Notice that for the partial integration steps we assumed $a>1$ to eliminate the boundary terms. Therefore this expression only holds for $a>1$. This is the relation we used in the main text of the paper.

\section{Summation of all ring  diagrams}
\label{sec:nonV}

We now derive the final formula for the correlation energy starting from the expression~\eqref{B}. If we insert in this expression our approximation for $Q$ and performing the angular $q$ integrals we obtain the following:
\begin{align}
\varepsilon^{c}_D \approx\delta_D-\dfrac{2D}{\pi^{3}}\int\limits_{-\infty}^{\infty}du \,R^2_{\frac{D-1}{2}}(u,\xi)\int\limits_{0}^{1}\dfrac{dq}{q^{D-2}}\sum\limits_{n=2}^{\infty}\frac{(-1)^{n}}{n}\left(\dfrac{4R_{\frac{D-1}{2}}(u,\xi)r_{s}}{\alpha_{D}\pi q^{D-1}}\right)^{n-2}.
\label{eqB1}
\end{align}
In this expression $\delta_D$ regulates the approximation such that the second order energy contribution is exact, namely,
\begin{multline}
\delta_D=\varepsilon^{(2)}_{c}+\dfrac{D}{\pi^{3}}\int\limits_{-\infty}^{\infty}[R_{\frac{D-1}{2}}(u,\xi)]^{2}du\int\limits_{0}^{1}\dfrac{dq}{q^{D-2}} \\
=\lim_{\beta \to 0}\left[-\left(\dfrac{\Gamma(\frac{D}{2}+1)\Gamma^{2}(\frac{D-1}{2})}{4\pi^{\frac{3D}{2}+1}}\right)\int\limits_{\beta}^{\infty}\dfrac{d^{D}q}{q^{2D-2}}\Xi_{1}^{(D)}(\textbf{q};k_{\uparrow},k_{\downarrow})
+\dfrac{D}{\pi^{2}}\int\limits_{\beta}^{1}\dfrac{dq}{q^{D-2}}\int\limits_{0}^{1}x[1-x^{2}]^{\frac{D-3}{2}}\int\limits_{0}^{1}y[1-y^{2}]^{\frac{D-3}{2}}\Xi_{2}^{(D)}(x,y;k_{\uparrow},k_{\downarrow})dx\,dy\right] \nonumber
\end{multline}
where $$\Xi_{1}^{(D)}(\textbf{q};k_{\uparrow},k_{\downarrow})=\frac{1}{4}\sum\limits_{i,j=\uparrow, \downarrow}\int\limits_{A_{ij}}\dfrac{d^{D}p_{1}d^{D}p_{2}}{q^{2}+\textbf{q}(\textbf{p}_{1}+\textbf{p}_{2})}, \,\,\,\,\,\,\,\,\,\Xi_{2}^{(D)}(x,y;k_{\uparrow},k_{\downarrow})=\frac{1}{4}\left[\frac{k_{\uparrow}^{2D-1}+k_{\downarrow}^{2D-1}}{x+y}+\frac{2(k_{\uparrow}k_{\downarrow})^{D-1}}{xk_{\uparrow}+yk_{\downarrow}}\right]$$ with $A_{ij}=(|\textbf{p}_{1}|<k_{i})\cap(|\textbf{p}_{1}+\textbf{q}|>k_{i})\cap(|\textbf{p}_{2}|<k_{j})\cap(|\textbf{p}_{2}+\textbf{q}|>k_{j})$.

To alleviate the notation  we define $\beta_{D}(u,\xi)\equiv\dfrac{4}{\alpha_{D}\pi}R_{\frac{D-1}{2}}(u,\xi)$. Substituting this in equation \ref{eqB1} we obtain:
\begin{align}
\varepsilon^{c}_D&\approx\delta_D-\dfrac{2D}{\pi^{3}}\int\limits_{-\infty}^{\infty}du\frac{[R_{\frac{D-1}{2}}(u,\xi)]^{2}}{(\beta_{D}(u,\xi)r_{s})^{2}}\left[\dfrac{\beta_{D}(u,\xi)r_{s}}{2}-\int\limits_{0}^{1}dq\,q^{D}\ln\left(1+\dfrac{\beta_{D}(u,\xi)r_{s}}{q^{D-1}}\right)\right]\\
&=\delta_D-\dfrac{2D}{\pi^{3}}\int\limits_{-\infty}^{\infty}du\frac{[R_{\frac{D-1}{2}}(u,\xi)]^{2}}{(\beta_{D}(u,\xi)r_{s})^{2}}\left[\dfrac{\beta_{D}(u,\xi)r_{s}}{2}-\int\limits_{0}^{1}\frac{dy}{D-1}y^{\frac{2}{D-1}}\ln\left(1+\dfrac{\beta_{D}(u,\xi)r_{s}}{y}\right)\right].
\label{b1}
\end{align}
We made the substitution $y=q^{D-1}$. Furthermore we define the quantity $\alpha\equiv\frac{2}{D-1}$. Unfortunately, since the latter is not in general an integer, we can not solve this integral in a closed form. To evaluate it we must expand the integrand in a Taylor series. To do this we must separate the integral into two parts: one part smaller than and one part greater than $\beta_{D}(u)r_{s}$. By doing so, one stays in the convergence radius of the corresponding series. Let us consider first the lower part of the interval, namely,
\begin{align}
\int\limits_{0}^{\beta_{D}(u,\xi)r_{s}}\frac{\alpha y^{\alpha}}{2}\ln\left(1+\frac{\beta_{D}(u,\xi)r_{s}}{y}\right)=\frac{\alpha}{2}\int\limits_{0}^{\beta_{D}(u,\xi)r_{s}}y^{\alpha}\ln\left(1+\frac{y}{\beta_{D}(u,\xi)r_{s}}\right)-\frac{\alpha}{2}\int\limits_{0}^{\beta_{D}(u,\xi)r_{s}}y^{\alpha}\ln\left(\frac{y}{\beta_{D}(u,\xi)r_{s}}\right)dy.
\end{align}
The second integral of the right hand side of this equation can be easily done, yielding $-[\beta_{D}(u,\xi)r_{s}]^{\alpha+1}/(\alpha+1)^{2}$. The first integral on the right hand can be Taylor expanded. After doing so, we arrive at:
\begin{align}
\int\limits_{0}^{\beta_{D}(u,\xi)r_{s}}y^{\alpha}\ln\left(1+\frac{y}{\beta_{D}(u,\xi)r_{s}}\right)=-\sum\limits_{n=1}^{\infty}\frac{(-1)^{n}}{n(\beta_{D}(u,\xi)r_{s})^{n}}\int\limits_{0}^{\beta_{D}(u,\xi)r_{s}}y^{n+\alpha}dy=-\sum\limits_{n=1}^{\infty}\frac{(-1)^{n}}{n}\frac{(\beta_{D}(u,\xi)r_{s})^{\alpha+1}}{n+\alpha+1}.
\end{align}
To perform the upper half of the integration interval, let us define $\delta_{a,b}$ as the Kronecker delta and a related quantity $\Delta_{a,b}=1-\delta_{a,b}$ which works as the complement of the Kronecker delta. With these definitions we can expand and eventually calculate the integral
\begin{align}
\int\limits_{\beta_{D}(u,\xi)r_{s}}^{1}\frac{\alpha y^{\alpha}}{2}\ln\left(1+\frac{\beta_{D}(u,\xi)r_{s}}{y}\right)&=-\frac{\alpha}{2}\sum\limits_{n=1}^{\infty}\frac{(-\beta_{D}(u,\xi)r_{s})^{n}}{n}\int\limits_{\beta_{D}(u,\xi)r_{s}}^{1}y^{\alpha-n}dy\\
=-\frac{\alpha}{2}\sum\limits_{n=1}^{\infty}\frac{\Delta_{n,\alpha+1}(-1)^{n}}{n}\frac{(\beta_{D}(u,\xi)r_{s})^{n}-(\beta_{D}(u,\xi)r_{s})^{\alpha+1}}{\alpha+1-n}&+\frac{\alpha}{2}\sum\limits_{n=1}^{\infty}\frac{\delta_{n,\alpha+1}(-1)^{n}}{n}(\beta_{D}(u,\xi)r_{s})^{n}\ln(\beta_{D}(u,\xi)r_{s}).
\end{align}
Let us define $\Sigma_{D}\equiv\sum\limits_{n=1}^{\infty}\left[\frac{(-1)^{n}}{n(n+\alpha+1)}-\frac{\Delta_{n,\alpha+1}(-1)^{n}}{n(\alpha+1-n)}\right]-\frac{1}{(\alpha+1)^{2}}$ and rewrite the content of the square brackets in Eq.~\eqref{b1}:
\begin{multline}
\dfrac{\beta_{D}(u,\xi)r_{s}}{2}-\int\limits_{0}^{1}\frac{dy}{D-1}y^{\frac{2}{D-1}}\ln\left(1+\dfrac{\beta_{D}(u,\xi)r_{s}}{y}\right)\\
=\frac{\beta_{D}(u,\xi)r_{s}}{2}+\frac{\alpha}{2}\Sigma_{D}(\beta_{D}(u,\xi)r_{s})^{\alpha+1}+\frac{\alpha}{2}\sum\limits_{n=1}^{\infty}\frac{\Delta_{n,\alpha+1}(-\beta_{D}(u,\xi)r_{s})^{n}}{n(\alpha+1-n)}-\frac{\alpha}{2}\sum\limits_{n=1}^{\infty}\frac{\delta_{n,\alpha+1}(-\beta_{D}(u,\xi)r_{s})^{n}}{n}\ln(\beta_{D}(u,\xi)r_{s}).
\label{b2}
\end{multline}
Now let us take a more closer look at $\alpha=\frac{2}{D-1}$. If we restrict our calculations to dimensions greater than $1$, we have that $\alpha\in (0,\infty)$. This means that $\alpha+1$ is never $1$, therefore we can drop the $n=1$ term of the second series on the right hand side of Eq.~\eqref{b2}. In contrast to this, the $n=1$ term of the first series is always there and cancels the first term of the right hand side in Eq.~\eqref{b2}. That allows us to simplify the above expression, and arrive at:
\begin{multline}
\dfrac{\beta_{D}(u,\xi)r_{s}}{2}-\int\limits_{0}^{1}\frac{dy}{D-1}y^{\frac{2}{D-1}}\ln\left(1+\dfrac{\beta_{D}(u,\xi)r_{s}}{y}\right)\\
=\frac{\alpha}{2}\Sigma_{D}(\beta_{D}(u,\xi)r_{s})^{\alpha+1}+\frac{\alpha}{2}\sum\limits_{n=2}^{\infty}\frac{(-\beta_{D}(u,\xi)r_{s})^{n}}{n}\left(\frac{\Delta_{n,\alpha+1}}{\alpha+1-n}-\delta_{n,\alpha+1}\ln(\beta_{D}(u,\xi)r_{s})\right).
\end{multline}
Since we are only interested in terms that do not vanish when $r_{s}$ approaches zero, we consider the expression $\frac{1}{(\beta_{D}(u,\xi)r_{s})^{2}}$ times the equation above. If we now look in the limit of small $r_s$ we can perform some simplifications: The first term vanishes for small radii if $\alpha >1$ which leads to the step function in the next expression. Furthermore, the parts of the series will all vanish for small radii if and only if $n>2$ which means that the series truncates at $n=2$. Since it started from $2$ the series has in this limit only one term, namely the $n=2$. Performing all these steps we get the following closed expression:
\begin{multline}
\frac{\alpha}{2}\Sigma_{D}(\beta_{D}(u,\xi)r_{s})^{\alpha-1}\theta(1-\alpha)+\frac{\alpha\Delta_{1,\alpha}}{4(\alpha-1)}-\frac{\alpha\delta_{1,\alpha}}{4}\ln(\beta_{D}(u,\xi)r_{s})\\
=\frac{\Sigma_{D}(\beta_{D}(u,\xi)r_{s})^{-\frac{D-3}{D-1}}}{D-1}\theta(D-3)-\frac{\Delta_{3,D}}{2(D-3)}-\frac{\delta_{3,D}}{4}\ln(\beta_{D}(u,\xi)r_{s}),
\end{multline}
where we shave substituted  the definition of $\alpha$. The full expression for the correlation energy is then:
\begin{align}
\nonumber
\varepsilon^{c}_D\approx\delta_D&-\dfrac{2D\Sigma_{D}}{\pi^{3}(D-1)}\left(\dfrac{\alpha_{D}\pi}{4r_{s}}\right)^{\frac{D-3}{D-1}}\int\limits_{-\infty}^{\infty}[R_{\frac{D-1}{2}}(u,\xi)]^{\frac{D+1}{D-1}}du\,\theta(D-3)\\
&+\int\limits_{-\infty}^{\infty}\left[\frac{D\Delta_{3,D}}{\pi^{3}(D-3)}+\frac{D\delta_{3,D}}{2\pi^{3}}\ln\left(\dfrac{4r_{s}}{\alpha_{D}\pi}R_{\frac{D-1}{2}}(u,\xi)\right)\right][R_{\frac{D-1}{2}}(u,\xi)]^{2}du.
\end{align}
From this expression one can learn directly the different behavior for $D<3$, $D=3$, and $D>3$.

\twocolumngrid

\bibliography{Refs2}

\end{document}